**astro-ph/0309762 v2,   revised 1 Mar 2007**

# Galactic mass distribution without dark matter or modified Newtonian mechanics

Kenneth F Nicholson, retired engineer
key words:  galaxies, rotation, mass distribution, dark matter

## Abstract

Given the dimensions (including thickness) of a galaxy and its rotation profile, a method is shown that finds the mass and density distribution in the defined envelope that will cause that rotation profile with near-exact speed matches.  Newton's law is unchanged.  Surface-light intensity and dark matter are not needed.   Results are presented in dimensionless plots allowing easy comparisons of galaxies.  As compared with the previous version of this paper the methods are the same, but some data are presented in better dimensionless parameters.  Also the part on thickness representation is simplified and extended, the contents are rearranged to have a logical buildup in the problem development, and more examples are added.

## 1.  Introduction

Methods used for finding mass distribution of a galaxy from rotation speeds have been mostly those using dark-matter spherical shells to make up the mass needed beyond the assumed loading used (as done by van Albada et al, 1985),  or those that modify Newton's law in a way to approximately match measured data  (for example MOND,  Milgrom, 1983).   However the use of the dark-matter spheres is not a correct application of Newton's law, and (excluding relativistic effects) modifications to Newton's law have not been justified.

In the applications of dark-matter spherical shells it is assumed that all galaxies are in two parts, a thin disk with an exponential SMD loading (a correct solution for rotation speeds by Freeman, 1970) and a series of spherical shells centered on the galaxy center.   The two constants needed for the exponential disk loading are supposed to be obtained from surface light intensity and a reasonable mass / light ratio.   However few galaxies if any have disks with an exponential loading like that, and the spherical shells had to be added to have enough mass in the outer radii to cause sufficient speeds to match experiment.  So the assumption that all galaxy disks must have that type of exponential loading is false, and so far there has been no evidence of the spherical shells (called dark matter) except that they can be used mathematically to get an approximate match for the measured rotation speeds.  The statement often made that the shells are necessary to satisfy the gravitational effects for galaxies is also false, as will be shown here.

Further problems are associated with the use of the added spherical shells.  Multiple solutions are possible for a given rotation profile, even allowing solutions with all spheres and no disk, but there must be only one mass distribution in a given galaxy geometry that can cause a given rotation profile. Also the attractions on a shell of the disk parts that are outside, and of disk parts and spheres inside, have not been included.  These effects would cause the shells to be pulled into the disk, or require orbiting components of the shells.  If there were such orbiting components, their effects when passing through the disk would be easily seen near our solar system.

Modifications to Newton's law have required an increase in strength of the gravity field as it gets weak with distance, an  effect that is not shown by experiment for anything that normally decreases with the square of the distance.  If it can be shown that Newton's law can be used correctly, to find, with only one possible solution, the mass distribution inside a galaxy's envelope that causes a measured rotation profile, there would be no basis for the use of extra matter outside a galaxy's envelope, or for the modification of Newton's law.  It could then be said with confidence (at least for galaxies) that dark matter doesn't exist, and that Newton's law (excluding relativistic effects) still works fine.

   In this paper Newton's law is unchanged, with each particle attracting every other, and all matter in a galaxy remains inside an envelope defined by its dimensions.  Like many other papers on this subject, it is assumed that the galaxies are axisymmetric and symmetric in z (normal to the disk), the rotation speeds are measured at the center plane, and there is a definite galaxy rim where the last speed reading is obtained.  Also it is assumed that there is solid rotation inside the most inner speed reading.  Actually, the speeds are probably measured at the surface, the galaxies are not often closely symmetric, and there may be some more matter outside the last reading, but these assumptions allow a reasonable solution. Only gravity effects are considered, with no gas pressure, shear force, electromagnetic, or relativistic effects.

## 2.  Method

   A galaxy is represented by a series of rings, each split into 720 constant-density (equivalent) segments represented by rods with axes normal to the galaxy plane.   Using an estimated density distribution in the z direction normal to the disk plane, an equivalent thickness is found that has a constant density in z. This constant-density equivalent thickness will cause a ring to have the same SMD and nearly the same gravitational effects as the one with the estimated density distribution.   With its equivalent thickness and a unit density, each ring i will cause an acceleration on a test mass at radius j  of amount a(i,j).  The total acceleration  A(j) at radius j , is then found by adding the effects of all rings with their appropriate densities.  This is the forward problem, finding the acceleration profile (and thus rotation speeds) caused by a given mass distribution.  The reverse problem, mass distribution from the rotation profile, can be done in a similar way, since all necessary data are known.  The matrix equations are:

$$a(i, j) \bullet \rho(i) = A(j) \quad , \quad \text{forward  problem} \qquad\qquad (1)$$

$$\rho(i) = a^{-1}(i, j) \bullet A(j) \quad , \quad \text{reverse  problem , where } a^{-1}(i, j) = \text{inverse of } a(i, j) \qquad (2)$$

   These equations show that for a given thickness and density distribution  there can be only one rotation profile.  Also for a given thickness and rotation profile, there can be only one density distribution.

## 3. Thickness Estimates

   Side views as in figure 1 are the best basis for thickness and density-pattern estimates.  Also a measured thickness of the Milky Way was shown in an article by Bok (1981).

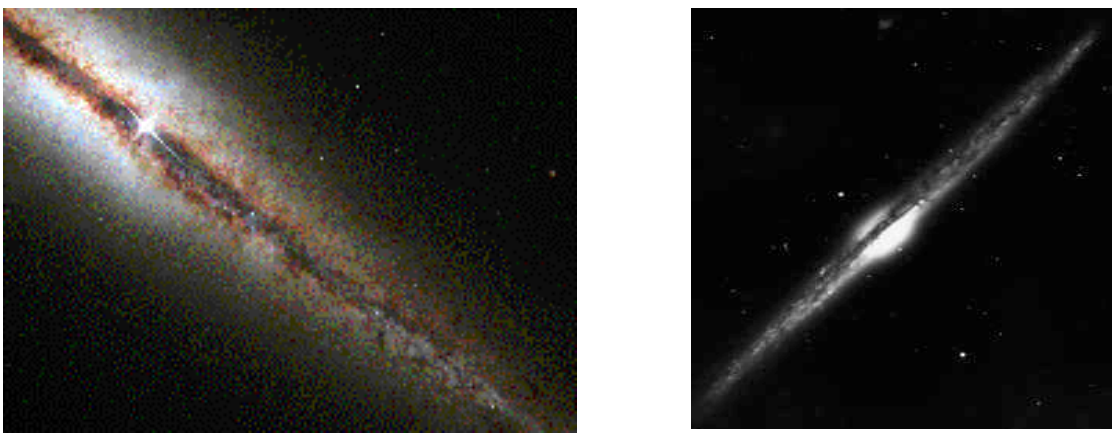

Fig. 1.  Side views of (left) NGC 4013 by Hubble, and  NGC 4565 from Jeff MacQuarrie's home page

The galaxies of figure 1 appear to have very similar thickness distributions for the large bodies near the mid plane, but NGC 4013 has an "atmosphere" of gas/dust and the other does not. The measured thickness of the Milky Way was somewhat thicker and seemed to be for the envelope of stars. The Milky Way also shows much gas/dust. The density variation in z appears to be about constant with radius for these examples. Not so for galaxies like the Sombrero galaxy, and for those the method shown below must be extended to account for changes with radius. Some ideas for this are at the end of this section.

To reduce computing time because of complications caused by an estimated density distribution model in z, normal to the disk, an equivalent segment is found that has the same mass, SMD, and gravity effects. This can be done by trial and error, computing the gravity effects of the chosen model for a single ring (of say 5% thickness) and repeating for different heights of the constant-density segment until a match is found. However a much faster method with acceptable accuracy is shown here.

Since a ring of a galaxy is represented by 720 rods (as segments) with axes normal to the disk, the quick method finds the maximum acceleration per unit mass caused by two rods for a test mass passing between the rods that have a model density distribution in z. After that is found for the model by computer integration, the maximum acceleration per unit mass is matched to find the equivalent rod length for the constant-density distribution.

For an arbitrary z distribution of density, dimensionless acceleration per unit mass is found by computer integration for several x, and a maximum found.

$$\frac{dA}{G} = \frac{2\, a\, rho\, dz\ x}{\left(x^2 + y^2 + z^2\right)^{1.5}} \quad , \quad \text{then} \quad Ad \equiv \frac{Ay^2}{2\, m\, G} \quad \text{dimensionless} \qquad (3)$$

for several x between 0.7 and 1.2 times y, to find a max

where  a = rod section area
m = mass of one rod
rho = local density
x = test mass distance perpendicular to rods
2y = rod separation, constant for a given ring
z = distance to dz from rod center

For a constant-density segment, the solution for Ad is

$$Ad = \frac{xd}{\left(1 + xd^2\right)\sqrt{1 + xd^2 + zdm^2}} \qquad (4)$$

Where  xd, zdm are  x/y and max z/y

So given max Ad for the model, and the optimum xd vs max Ad for the constant-density segments from the collected results in figure 2, a solution is found for zdm giving the height of the equivalent constant-density segment. From figure 2,

Admax < 0.118, xdopt = 0.98
0.118 < Admax < 0.247, xdopt = 0.98 - 1.209 (Admax-0.118)
0.247 < Admax, xdopt = 0.824 - 0.754 (Admax - 0.247)

so   $$zdm = \left[\left(\frac{xd}{\left(1+xd^2\right)Ad\max}\right)^2 - \left(1+xd^2\right)\right]^{0.5}$$   , where $xd = xdopt$   (5)

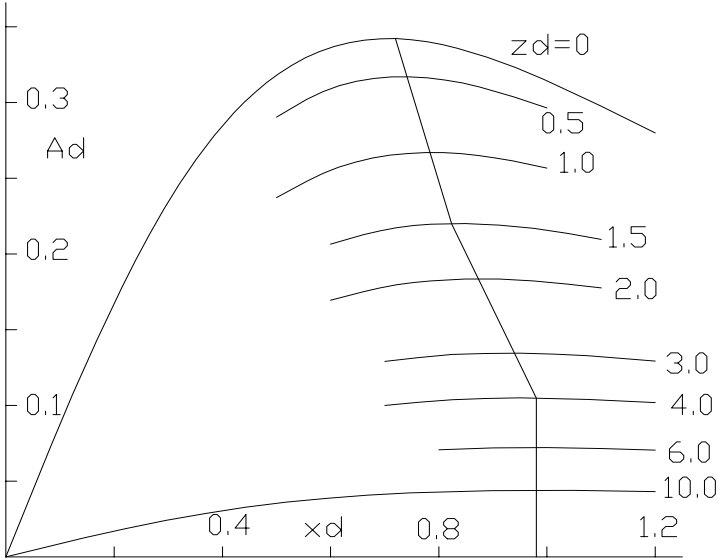

Fig 2. Dimensionless acceleration for a test mass passed
between two rods of constant density in z

Equivalent segments for several density distributions are in figure 3. The large-body thickness is the same for all, and the equivalent segments have the same mass as their models.

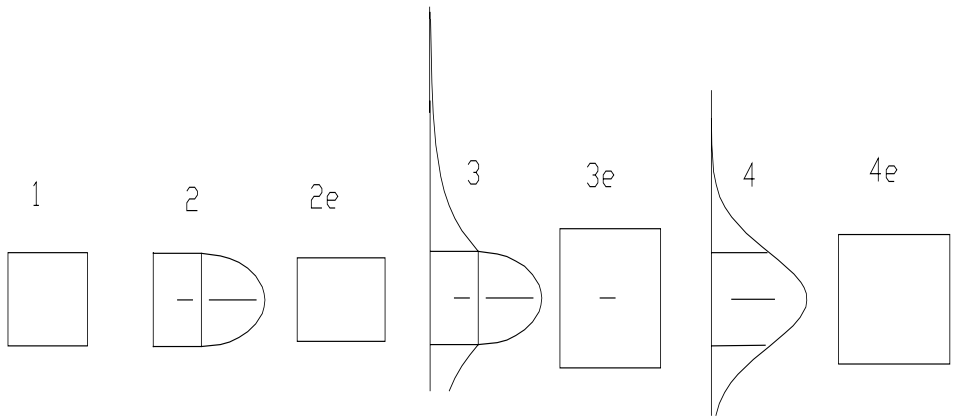

Fig 3.  Some reasonable density distributions with equivalent segments of  thickness  k  x  hb

1.  normal constant-density section with k=1, and thickness of the large bodies hb

2.  1/2 gas/dust with constant density, 1/2 large bodies with  elliptic distribution, k = 0.9066

3.  1/3 gas/dust constant density, 1/3 gas/dust in an exponential "atmosphere," 1/3 large bodies
    with elliptical distribution,  k = 1.3957

4.   error function distribution with 2/3 of mass inside hb,  k = 1.3261

Large-body thickness distributions used here are shown in figures 4 a,b,c. They seem adequate for most galaxies that are nearly symmetric disks. The last has no bulge and is much thicker, as implied by the results for NGC 6822 where the SMD does not show a strong increase near the center, and density seems far too high. Rd and hbd are the radius and large-body thickness divided by the maximum radius.

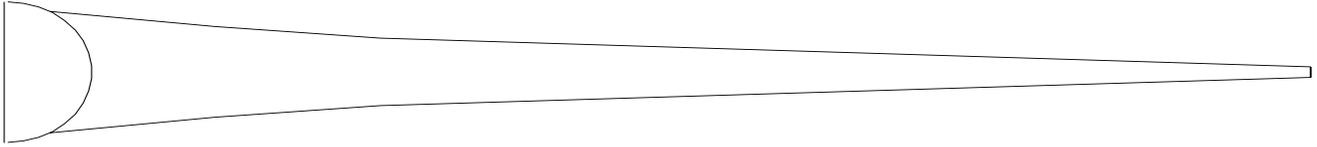

hbd = 0.1288 sqr(1 − (rd/0.0644)^2) , for rd < 0.0322 ,

then    hbd = 0..11158 exp ( − 2.315 (rd − 0.0322)) , for rd < 0.2857 ,

then    hbd = 0.062 − 0.0727 (rd − 0.28577)

Fig. 4a. Thickness distribution from picture in Milky Way article (Bok, 1981)

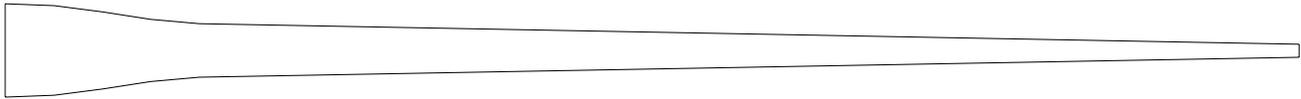

straight line between points
rd = 0, 0.05, 0.10, 0.15, 1.00
hbd = 0.07, 0.065, 0.05, 0.04, 0.01

Fig 4b. Thickness distribution estimated from picture of NGC 4013 (Fig 1.)

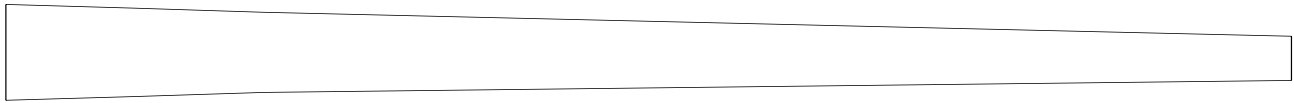

straight line between points
rd = 0, 0.1, 0.2, 1.0
hbd = 0.06, 0.055, 0.05, 0.02

Fig 4c. Thickness distribution estimated for small galaxies without significant bulge

Figure 5 shows some of the difficulties encountered with unusual galaxies. On the plus side there appears to be a definite rim. On the minus side the speeds would be hard to measure inside the half radius, and from this picture, it is hard to see if there is any large-body disk there. For the region inside the half radius it appears that light pressure from the central bulge has driven the large bodies out radially in the disk plane, and the gas/dust radially away from the bulge in all directions. So the gas/dust halo could be continuous from the bulge or even have a large empty region near the bulge. For any case the galaxy can again be represented by rings, but the density distribution might be discontinuous in z. Even so, equivalent sections could be found but they would have to be functions of radius. I have not tried to make the estimate.

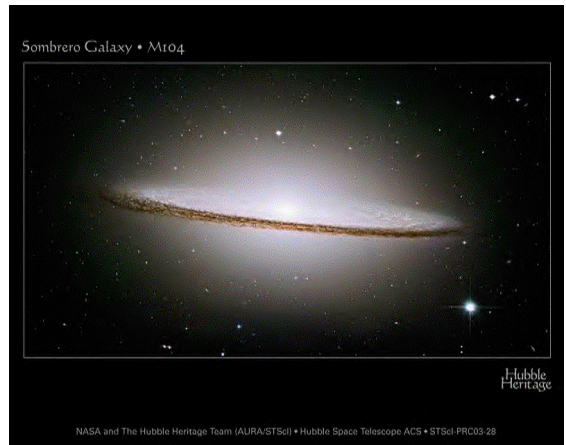

Fig. 5.   Sombrero galaxy M104, from Hubble

## 4.  Equation Development

The galactic disk is represented by rings (usually 40 to 80) each made up of 720 constant-denstity rods of length equal to the local equivalent thickness.  Each rod has the mass of the local incremental volume, and passes through its center of mass in a direction normal to the galactic plane. These rods allow an analytic result for the attraction of a single rod on a test mass, and because the test mass is not allowed to touch any rod, division by zero is avoided.  The attraction between a rod and a test mass is always correct.

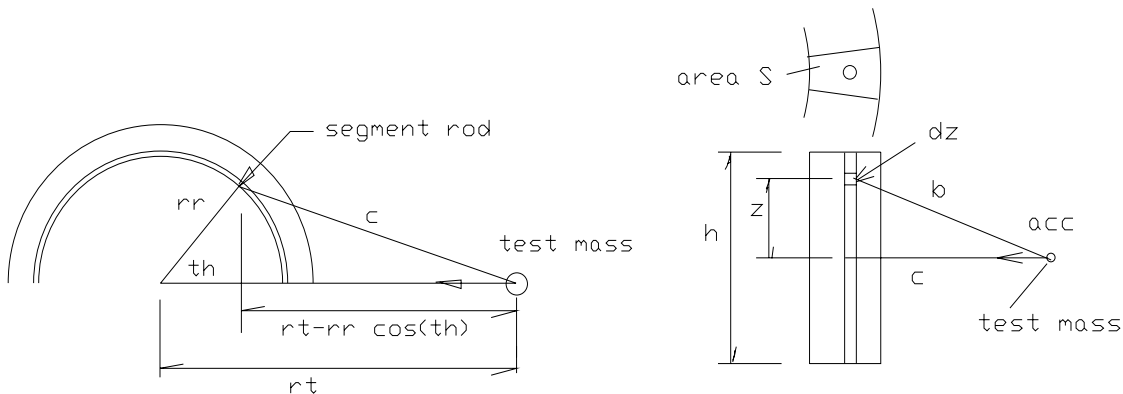

Fig 6.   Rod and galactic geometries

The acceleration caused by and toward a single rod is found by integration  to be:

$$acc = \frac{rho \ G \ dvol}{c \ \sqrt{c^2 + (h/2)^2}} \qquad (6)$$

Noting that $dvol = h \ r \ dth \ dr$ ,  the integral for $v^2$  at radius rt becomes:

$$v^2 = (pytks)^2 \times rt \times \int_0^{r\,max} rho \int_0^{2\pi} \frac{G \ dvol}{c\sqrt{c^2 + (h/2)^2}} \ \frac{(rt - rr \cos(th))}{c} \qquad (7)$$

In digital form, rearranged slightly for convenience, the equation for the forward problem is:

$$v^2(j) = (pytks)^2 \times rt(j) \times \sum_{i=1}^{Nr} rho(i) \left[ 2 \sum_{1}^{360} \frac{G \ dvol}{\sqrt{c^2 + (h/2)^2}} \times \frac{(rt(j) - rr \ cos(th))}{c^2} \right]_i \qquad (8)$$

where $c^2 = (rr \ sin(th))^2 + (rt - rr \ cos(th))^2$, $pc^2$, note that c is never zero

dvol = volume of the fundamental segment, $h \ r \ dth \ dr$, $pc^3$
dth = 0.5 degree
G = gravity constant, 4.498E-15 $pc^3 / (msuns / yr^2)$
h = galaxy equivalent thickness at radius r, pcs
Nr = number of rings
pytks = 9.778E5 (kms/sec)/(pcs/yr), changes speeds from pcs/yr to kms/sec
r = radius to centerline of ring, pcs
rho = equivalent density, msuns/$pc^3$
rm = radius to outer edge of ring, where speeds are measured, computed, and compared, pcs
rr = radius to rod used to represent fundamental segment mass, pcs
= (rm -dr) +dr/2 (rm-dr/3)/(rm-dr/2)
rt = radius to test mass, pcs
th = (i-1/2) dth, for i = 1 to 360 , degs
v = orbital speed of test mass at radius rt, kms/sec

To compare with the matrix form (1), it is seen that the summation inside the large brackets of (8) defines a(i,j), the acceleration of a test mass at radius j caused by the ring i with unit density. Thus, except that it uses the acceleration to solve for $v^2$, (8) is the same as the matrix equation (1).

Computing is done in dimensioned form and the results made dimensionless for plotting, as follows:

Ad = acceleration / (vkep$^2$/rmax), where vkep$^2$/rmax = the Keplerian acceleration at the rim
hd = h / rmax
md = m / mtot
rd = r / rmax
rhod = rho / rhoav , where rhoav = mtot /(equivalent volume)
SMDd = SMD / SMDav, where SMDav = mtot / ($\pi$ rmax$^2$ )
vd = v / vkep , where vkep = G mtot / rmax , the Keplerian speed at the rim

For the reverse problem, solutions for density and surface-mass density can be obtained by standard matrix methods (for example W H Press et al, 1987), or by iteration of the densities to obtain a match of the computed and measured speeds (Nicholson, 2000).

## 5. Results with the forward problem

To check the equations and coding and get familiar with the problem, comparisons were made with known analytic solutions, and trials done with some galaxy-like shapes. The first is the result for a single ring. The analytic solution for a zero diameter "wire" ring shows that the acceleration becomes infinite when the test mass is at z = 0 and at the ring, because the ring has infinite density. Yet a simple thought experiment shows that, for a "cloud-like" ring with thickness as in a galaxy, a test mass acceleration would be zero at the center of the galaxy, increase with radius going toward the ring, then pass smoothly through zero, changing sign as it is pulled inward toward the ring. Exactly at the ring, the acceleration should theoretically match the Keplerian value.

Figure 7 shows the computed result for a 5% thick single ring, matching "wire-ring" theory closely until near the rim, then doing as expected as the test mass goes through the ring.  The acceleration is positive toward the center to avoid minus signs in the computing.  Keplerian acceleration occurs slightly outside the ring, but the location shown is very close.  The dimensionless theoretical result for a wire ring is in (9), with the ring location at rd = 1.    As the ring gets thinner, say 1%, the results look more like theory for the wire ring, as expected.

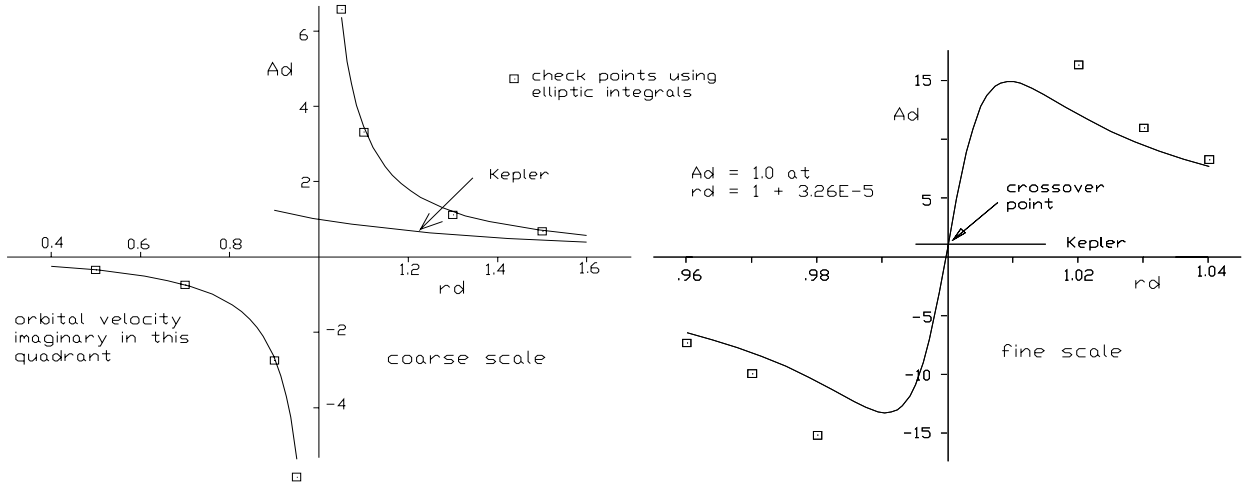

Fig. 7.  Test-mass acceleration passing through a 5% thick single ring

$$Ad = \frac{1}{\pi \, rtd} \left[ \frac{K(k)}{(1+rtd)} - \frac{E(k)}{(1-rtd)} \right], \ \text{with} \quad k^2 = \frac{4 rtd}{(1+rtd)^2} \tag{9}$$

"wire" ring acceleration

An often-checked analytic solution for an exponential loading of  SMD was found by Freeman (1970).  It uses zero thickness and requires the radius go to infinity.  I used 1% of rmax for thickness with a definite rim.  Freeman's result has been used for the disk portion of most papers using dark matter spheres.   A comparison  with this result is in figure 8.  A small difference in speeds shows up at the rim, where my mass stops and Freeman's keeps going on to infinite radius.  Thus near the rim my solution does not have the mass outside the rim attracting the test mass, and my speed there is greater.  The Freeman equations (from Binney and Tremaine, 1987) are:

For SMD = $\Sigma = \Sigma_o \, e^{-2y}$  ,   the solution is:

$$v^2(r) = 4 \, \pi \, G \, \Sigma_o \, R_d \, y^2 \, [ \, I_o(y) \, K_o(y) - I_1(y) \, K_1(y)] \tag{10}$$

$$m(r) = 2\pi \sum\nolimits_0 \, R_d^{\,2} \left[ 1 - \exp\left(-r/R_d\right)(1+\frac{r}{R_d}) \right], \quad \text{the mass inside r} \tag{11}$$

where y ≡ r / (2 $R_d$),  and I and K are modified Bessel functions of the first and second kind

note that $R_d$  is not the maximum galaxy radius here, but merely a shaping constant

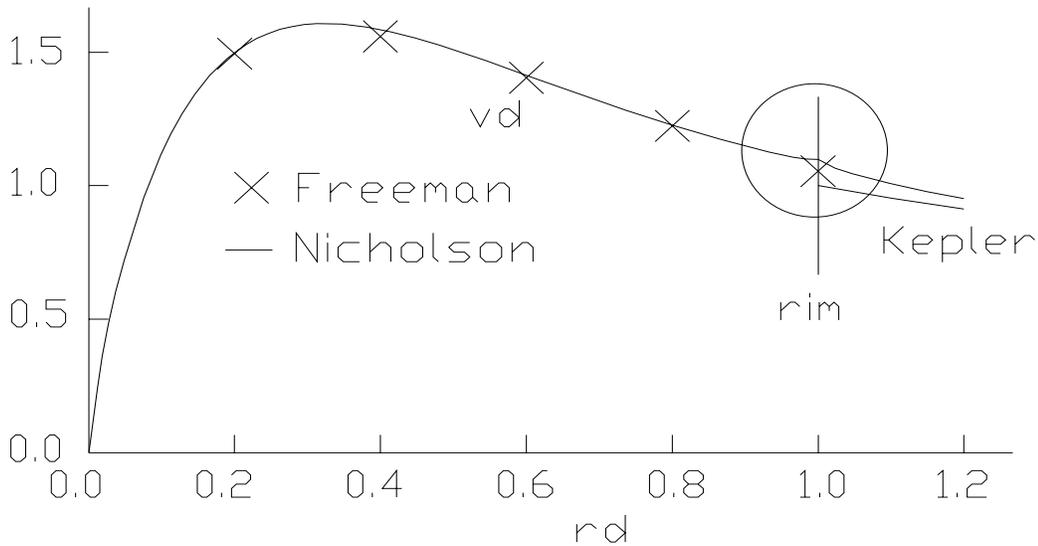

Fig. 8.  Comparison of rotation speeds for exponential loading

Trials were also done with constant-density sphere and disk, and arbitrary galaxy-like shapes as in figure 9.

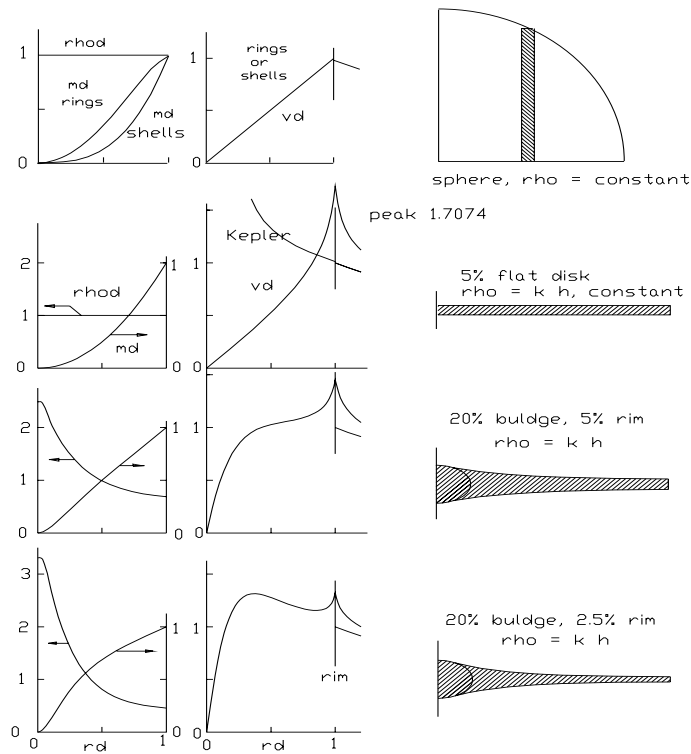

Fig. 9.   Trials shapes with the forward problem

The sphere shows that both using rings as done here, or spherical shells as in a textbook method, the same rotation profile results, although with rings, mass builds up much faster, as expected.  The high rim speed for the flat disk was a surprise.  The acceleration at the rim is almost three times that of the sphere with the same mass.  This result is apparently not well known, or Keplerian rotation profiles for galaxies would not have been expected when rotation speeds were first measured.  The constant-thickness disk, with the two shapes below that have density increasing with thickness, show how the rotation profile shifts toward the Keplerian as the mass moves to the center.  If the bottom two are averaged the resulting rotation profile would be almost flat, as in many galaxies.  The sharp increase at the rim is caused by the blunt edges that do not have decreasing density.

## 6.  Results with the reverse problem

In figure 10 an arbitrary size and rotation profile are used to show the effects of different disk thicknesses, and to show why the exponential loading causes such low rotation speeds except near the galaxy center.  Maximum radius is 20000 pcs, and the rotation profile is v = 200 (1− exp(−r/2000)) kms/sec.   Both thick and thin use the thickness distribution of figure 4b, but multiplied by 2.0  and 0.5 respectively.  So there is a 4 to 1 difference in thickness.

An exponential form for the SMD would have the shape ln(SMDd) = C − k  rd, as shown by the straight line on the plot.  This shows the discrepancy in mass for such a loading, and it is clear that the speeds resulting would be far too low.   Because of the way the results are made dimensionless the two thicknesses show essentially no differences on the plots, except at the bulge.  There ln(SMDd) and ln(rhod) increase with thickness, and SMDmax increases, but rhomax decreases.  In  general the changes caused by the increase in thickness are small for total mass, SMD, and  Kepler speed at the rim, and large for density and  total volume .

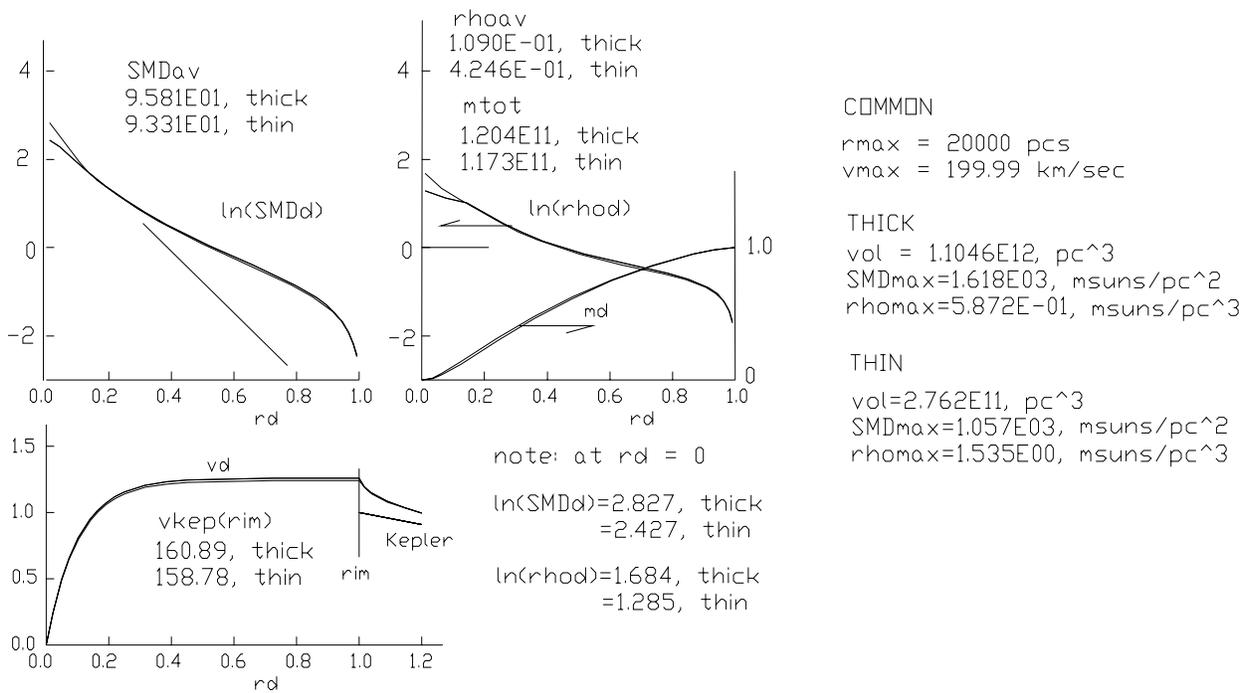

Fig. 10.  Imaginary galaxy  with 2 different thickness distributions, 60 rings

A real galaxy (figure 11) with a similar rotation profile is NGC 3198, used by van Albada et al (1985) for the basic paper behind dark matter.   The data are taken directly from their paper and extra points interpolated to allow 44 rings with even spacing.  It is assumed that a definite rim exists at the last speed reading. As compared with figure 10 the SMD increases more rapidly approaching the center of the galaxy, and this causes the steeper slope of  vd there..  In all of these plots the input rotation speeds are matched exactly for plotting purposes.  From the pictures of the galaxy I could detect no "atmosphere."  A large-body thickness distribution of figure 4b  was chosen with density distribution 2, giving an equivalent thickness of  0.9066 hb.

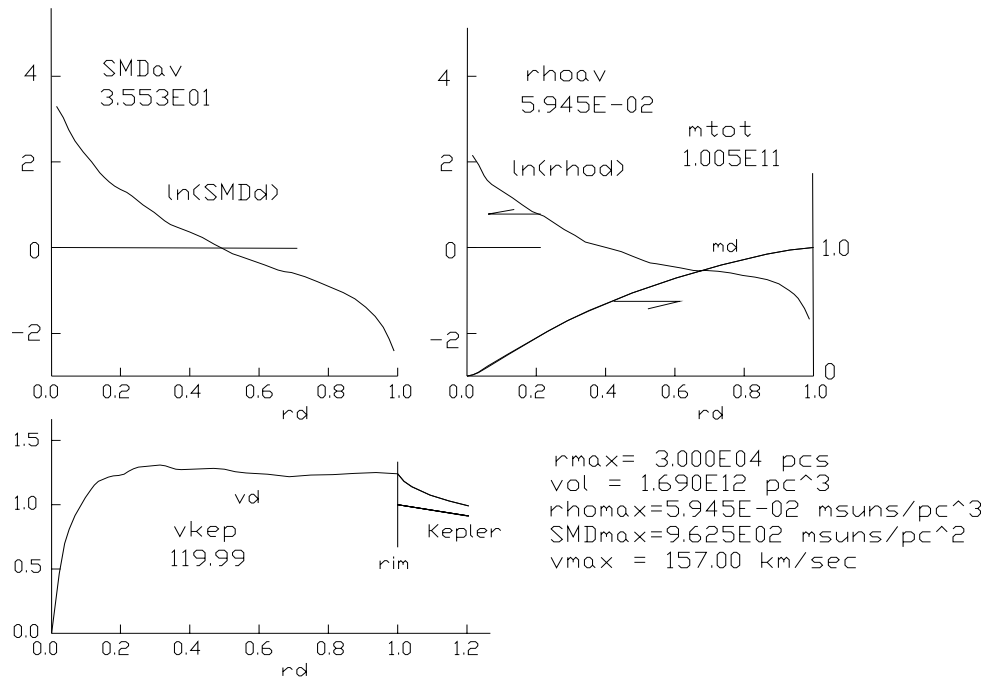

Fig. 11.  NGC 3198, data from van Albada et al (1985),
44 rings, thickness of figure 4b, density distribution 2e

My total mass of about 1.0 E11 msuns compares with 1.5 E11 for that of van Albada et al, when he used the minimum allowable fraction of spherical shells.  Using larger fractions for the shells, his total mass increases

UGC 9133 in figure 12 is a very large galaxy  (rmax = 102500 pcs ) with a very steep slope of measured speed  near the origin.  Even so the plots are similar, but distorted, as compared with figure 10.  Again the thickness used was from figure 4b  with density distribution 2e giving an equivalent thickness of 0.9066 hb.   This  was chosen but with no strong reason, except I could see no evidence of a gas / dust atmosphere in the picture.

Noordermeer et al  (2004) made some very good measurements of the rotation rates and tried to establish the mass distribution using several combinations of disk and dark-matter spherical shells, but could get no definite answers.  The method used here gives one definite and repeatable answer.

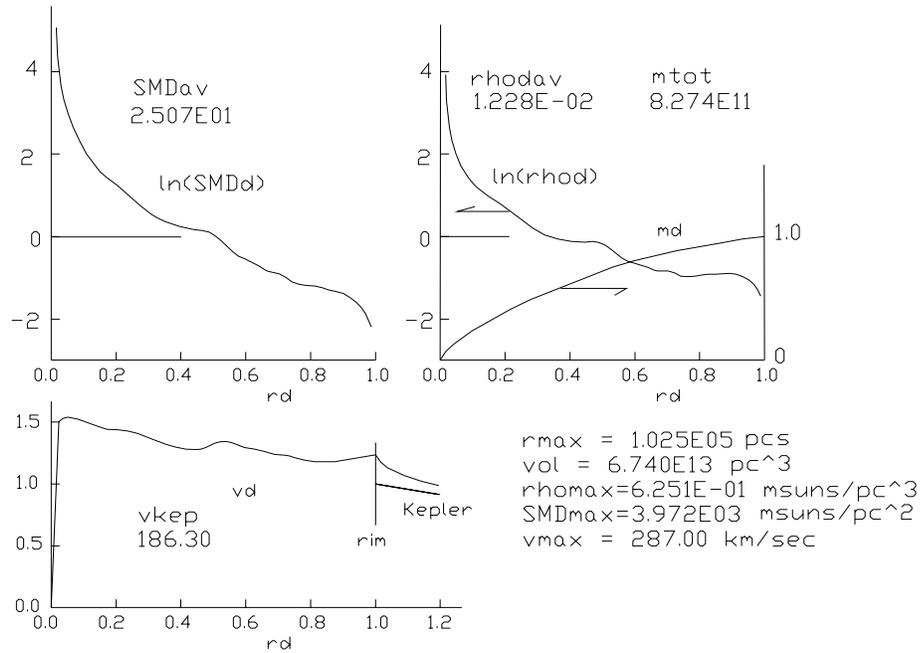

Fig. 12.  UGC 9133, data from Noordermeer et al (2004), 41 rings,
thickness of  figure 4b, density distribution 2e

The next example in figure 13 uses the speeds shown in Bok's article (1981).  The speeds were measured by Schmidt (1966) and Blitz (1980) and I could get no references.  The speeds may now be out of date.  However they serve as a suitable case that shows some extreme relative SMD's and densities at the center.  Thickness d was chosen because some Milky Way pictures show an "atmosphere ." The sharp rise in speed at the center of the Milky Way indicates "solid-body" rotation

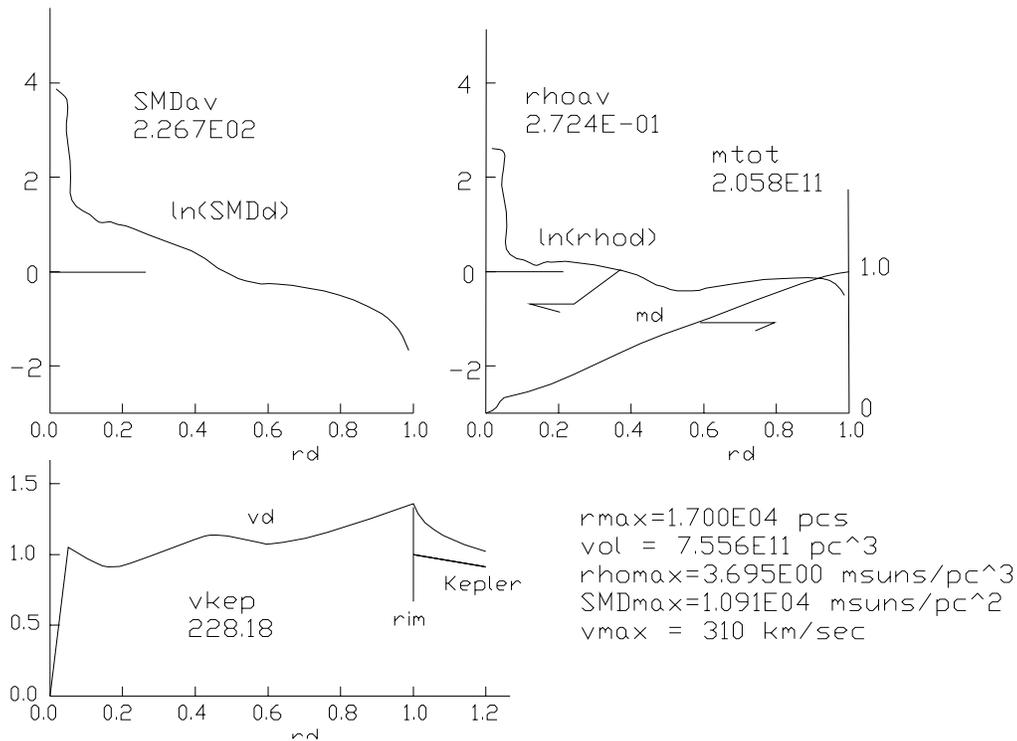

Fig. 13.  Milky Way, using speeds from Bok article (1981), 40 rings,
thickness of  figure 4a, density distribution 3e

The Milky Way is different from the previous examples in that it has a rising rotation speed with near-constant density in the outer half of the radius, and shows sudden jumps in SMD and density near the center. SMD jumps from 9.901E02 to 1.091E04 msuns/pc^2, and the density from 4.015E-01 to 3.695E00 msuns/pc^3, both about a factor of ten. It has been suggested that this might indicate a black hole. However the max density shows 3.695^(1/3) = 1.92 msuns/pc or about 0.52 pcs = 1.7 light years between sun-sized stars. This seems adequate to avoid many impacts, and with larger stars on the average there would be more clearance. So these results suggest the center is made up of a collection of stars.

At our Earth location (about at rd=0.5126) the equivalent density computed (a reasonable average) is 0.1843 msuns/pc$^3$. Binney and Tremaine, page 16, show an average of 0.18 , a nice check.. However this certainly doesn't show that good density accuracy applies to other galaxies, since those thicknesses are only estimates.

The Large Magellanic Cloud in figure 14 shows some drastic speed changes with radius. These speeds are from Alves et al (2000) but were actually measured by Kim (1998) and Kunkel et al (1999). Kim measured H1 speeds at radii less than 3100 pcs and Kunkel used carbon stars for radii more than 3000. I chose the H1 measurements at the one point where they met and disagreed, because they were more continuous. Like the speeds, the SMD and density results are quite lumpy. This demonstrates the ability of the program to handle large speed changes over a short radius change.

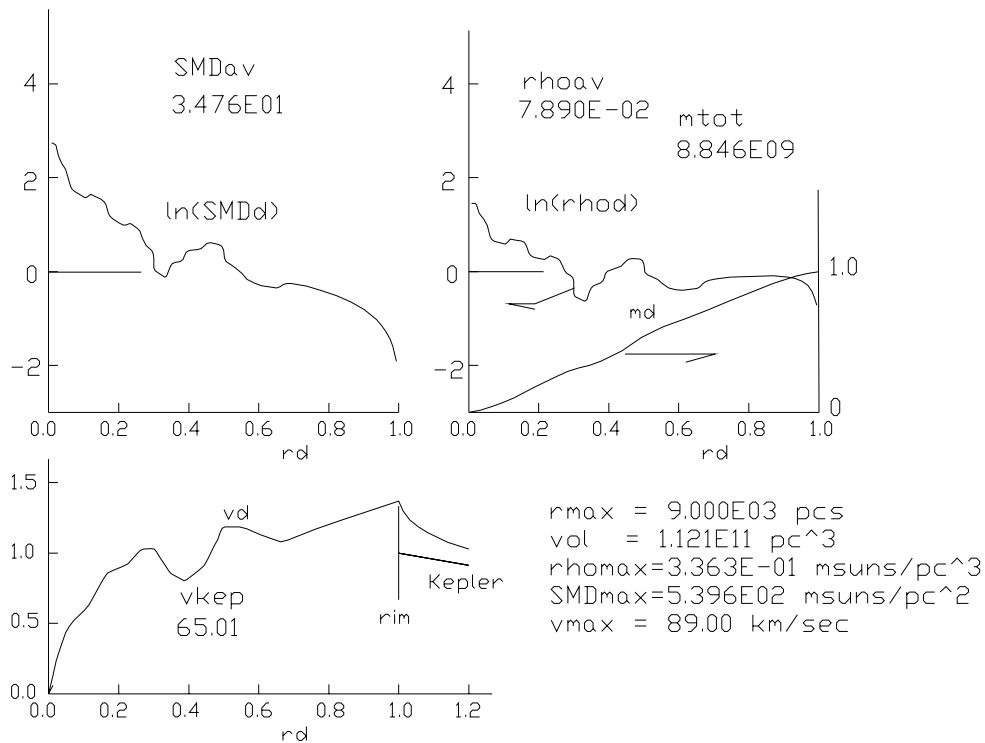

Fig. 14. Large Magellanic Cloud, speeds by Kim et al (1998) and Kunkel et al (1999)
80 rings thickness distribution of figure 4a, density distribution 3e

NGC 6822, the next and last example, is a small galaxy with rotation speeds well defined. The authors have found the rotation speeds very well but could not get a good answer for mass distribution, using many attempts with different combinations of disks and dark-matter spheres. They have defined many more than 100 rotation speeds, and the challenge is to match these speeds as well as possible with an appropriate mass loading. I did not try to match every speed but approximated the rotation curve with straight lines that show most of the bumps, and used 80 rings with thickness from figure 4c and density distribution 2e.

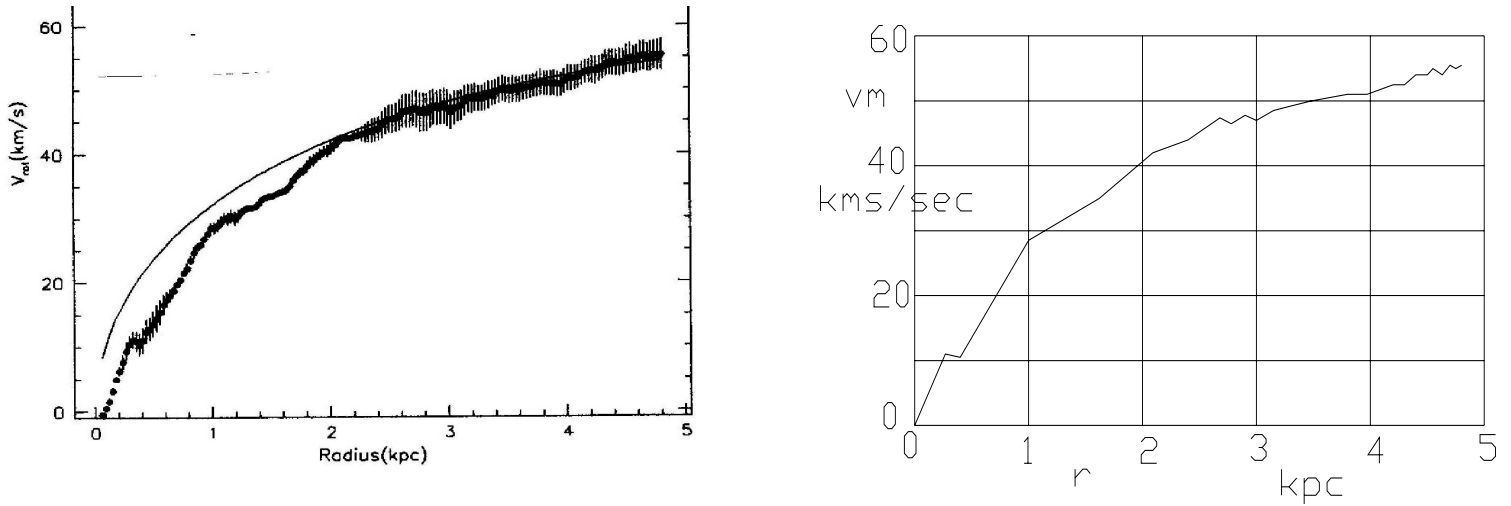

Fig. 14.  Rotation speeds for NGC 6822 from Weldrake et al (2002), and an approximation

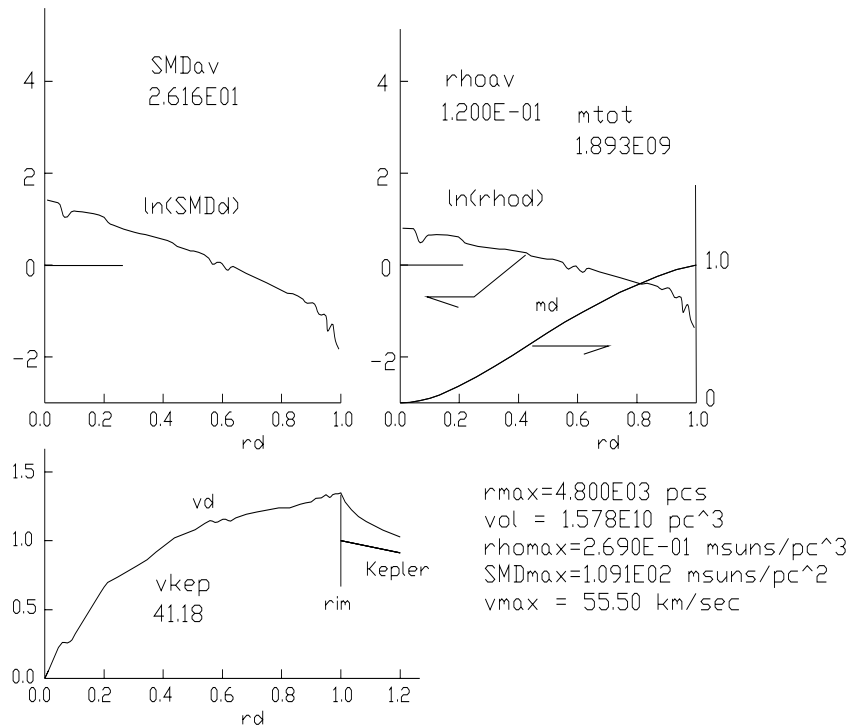

Fig. 15.   Results for NGC6822, 80 rings,  speeds from Weldrake et al (2002),
thickness figure 4c, density distribution 3e

The results in figure 15 show that the speeds of the approximation were matched  well.  Also the curves for SMD and rho indicate the center does not have a high concentration of mass, as in the other examples, and thus no bulge is assumed.   The shape of the SMD curve, with no strong buildup near the center, suggests this galaxy is young.   As energy is lost through friction, the galaxy mass should tend to concentrate in the center, and the rotation profile should tend toward solid rotation with rising speeds toward the rim.  This galaxy has rising speeds but no concentration of mass at the center.

A review of the results shows there is much more to be learned.   The table below shows some results in order of decreasing galaxy radius.

|  | rmax E-04 | vol E-11 | mtot E-11 | SMDav E-01 | rhoav E02 |
|---|---|---|---|---|---|
| UGC 9133 | 10.250 | 674.000 | 8.274 | 2.261 | 1.228 |
| NGC 3198 | 3.000 | 16.897 | 1.005 | 3.553 | 5.945 |
| Milky Way | 1.700 | 7.556 | 2.053 | 22.669 | 2.717 |
| L M Cloud | 0.900 | 1.121 | 0.089 | 3.476 | 7.890 |
| NGC 6822 | 0.480 | 0.158 | 0.019 | 2.616 | 12.001 |

The average densities of the Large Magellanic Cloud and NGC6822 are much higher than that of the Milky Way, but the SMD of the Milky Way is much larger than for those two, or for any of the others. This suggests that things do not scale in a simple way, and has caused a continuous search for errors in the equations and coding. So far the computing looks correct. Until this is checked further it seems fruitless to speculate on the cause.

## 5. Acknowledgements

## 6. Conclusions

1. Mass distribution can be found from galactic rotation speeds using Newton's law without modification, without any knowledge of surface-light intensity, and with no need for additional matter outside the galactic envelope.

2. Although thickness cannot often be measured, an equivalent thickness (accounting for an atmosphere) should be estimated and used as a way to improve the results for SMD, and to get an estimate for the density distribution.

3. Galaxies need no stars so far as dynamics are concerned, and there may be many galaxies that are so far undetectable, except perhaps by gravity effects. However there is no reason to say they would be made up of anything but ordinary matter (baryonic).

4. At least for galaxies, all matter is ordinary and a special kind of dark matter does not exist.